\documentclass[%
 reprint,
superscriptaddress,
 amsmath,amssymb,
 aps,pra
]{revtex4-2}

\usepackage{graphicx}
\usepackage[dvipsnames]{xcolor}
\usepackage{bm}
\usepackage{physics}
\usepackage{mathtools}
\usepackage{hyperref}
\usepackage{bbm}
\hypersetup{
    colorlinks,
    citecolor=blue,
    filecolor=blue,
    linkcolor=blue,
    urlcolor=blue
}
\usepackage{subcaption} 

\DeclareMathOperator*{\argmax}{argmax}

\definecolor{customOrange}{RGB}{253,164,64}
\definecolor{customBlue}{RGB}{76,89,154}

\usepackage{tikz}

\makeatletter

\newcommand{\Rmnum}[1]{\expandafter\@slowromancap\romannumeral #1@}
\makeatother

\begin{document}

\title{Even-parity precession protocol for detecting nonclassicality and entanglement}

\author{Jinyan Chen}
\affiliation{Centre for Quantum Technologies, National University of Singapore, 3 Science Drive 2, Singapore 117543}

\author{Jackson Tiong}
\affiliation{Department of Physics, National University of Singapore, 2 Science Drive 3, Singapore 117542}

\author{Lin Htoo Zaw}
\affiliation{Centre for Quantum Technologies, National University of Singapore, 3 Science Drive 2, Singapore 117543}

\author{Valerio Scarani}
\affiliation{Centre for Quantum Technologies, National University of Singapore, 3 Science Drive 2, Singapore 117543}
\affiliation{Department of Physics, National University of Singapore, 2 Science Drive 3, Singapore 117542}

\date{\today}

\begin{abstract}
We introduce an even-parity precession protocol that can detect the nonclassicality of some quantum states using only measurements of a uniformly-precessing variable at different points in time. Depending on the system under study, the protocol may detect the Wigner negativity of a single quantum harmonic oscillator or of a single spin $j\geq 2$; the non-Gaussian entanglement of two harmonic oscillators; or genuine multipartite entanglement of a spin ensemble, whose total spin is an integer. Unlike in other nonclassicality tests, simultaneous or sequential measurements are not required. Our protocol can also detect states that commute with the parity operator, which were missed by similar protocols built from Tsirelson's original precession protocol. This work also closes a long-standing gap by showing the possibility of detecting the Greenberger-Horne-Zeilinger entanglement of an even number of qubits using only collective spin measurements.
\end{abstract}

\maketitle


\section{\label{sec:level1}Introduction}
Uniform precession is found in the dynamics of many physical systems, from the orbits of large celestial bodies, to the harmonic motion of mesoscopic objects in optomechanical traps, and the rotation of microscopic spins in a uniform magnetic field. As a uniformly-precessing observable has the same dynamics in both classical and quantum theory, one would not expect the simple observation of its values at different times to reveal any nonclassical signatures. Yet, in an unpublished preprint \cite{tsirelson-og} Tsirelson showed this expectation to be wrong for the harmonic oscillator, whose dynamics is a precession in phase space. He proved that the probability of finding the oscillator with position $x>0$ at one out of $K=3$ times exceeds the maximal value allowed by classical mechanics for some suitable quantum states. This is an example of a mechanical task with a demonstrable quantum advantage, in the same vein as the probability backflow of a freely-evolving particle \cite{quantum-backflow} and the enhanced distance traveled by quantum projectiles \cite{quantum-rockets}. In these tasks, simultaneous or sequential measurements are not required, in contrast to other single-system nonclassicality tests, like contextuality and Leggett-Garg inequalities \cite{contextuality-review,LG-review}. 

Recently, Tsirelson's original protocol was extended to what we will call \emph{odd-parity precession protocols}, which certify nonclassicality (in the form of Wigner negativity \cite{wigner-function-review}) by probing a uniformly-precessing observable at one of $K$ suitably chosen times, where $K$ is odd \cite{tsirelson-spin}. In the same work, It was also noticed that the precession does not need to be in phase space: it could be in real space, if it involves non-commuting observables. Thus, a precession of position (circular orbit) cannot show any non-classicality, but a precession of angular momentum can, and actually does. This led to the extension of the Tsirelson idea to discrete variables, specifically to the precession of spins.

The odd-parity precession protocols were further proved to be able to witness entanglement when applied to collective degrees of freedom. Notably, some non-Gaussian entangled states of two harmonic oscillators can be detected using only measurements of their center-of-mass position, and without the risk of the false positives typical of witnesses based on uncertainty relations \cite{tsirelson-harmonic-entanglement}. In spin ensembles, genuine multipartite entanglement can be detected measuring only total angular momentum. While similar witnesses of this type were known, this is the first family that also detects Greenberger-Horne-Zeilinger (GHZ) states \cite{tsirelson-spin-entanglement}.

Like any linear non-classicality witness, odd-parity precession protocols cannot detect all states of interest (be they Wigner-negative or entangled). In particular, by reason of their symmetry, they miss all states that commute with the parity operator, as well as even-number GHZ states. In this paper, we introduce a family of \emph{even}-parity precession protocols. They also require only measurements of a uniformly-precessing observable at one of $K$ suitably chosen times, but $K$ is now even. These protocols can be implemented in the same way as the odd-parity ones, but requires a different expression for the score, as checking for positivity of the observable leads to trivial scores by symmetry. We show that these protocols detect Wigner negativity and entanglement in close analogy to the odd-parity protocols for a different set of states. In particular, when it comes to spin ensembles, we close the problem of detecting GHZ states using collective spin measurements by exhibiting a protocol that does it for an even number of spins (the case of an odd number of spins having been solved in \cite{tsirelson-spin-entanglement}).

In Sec.~\ref{sec:classical} we introduce the even-parity precession protocols and show the upper bound of the classical system. In Sec.~\ref{sec:quantum}, we discuss the two quantum violations, the quantum harmonic oscillator in sec. \ref{sec:QHO1} and spin angular momentum in sec. \ref{sec:SAM1}. In Sec.~\ref{sec:entanglement}, we demonstrate that the protocols can be used to detect entanglement when measuring collective coordinates. 

\section{\label{sec:classical}The even-parity precession protocols and their classical bound}

\begin{figure*}
    \centering\includegraphics[width = 170mm]{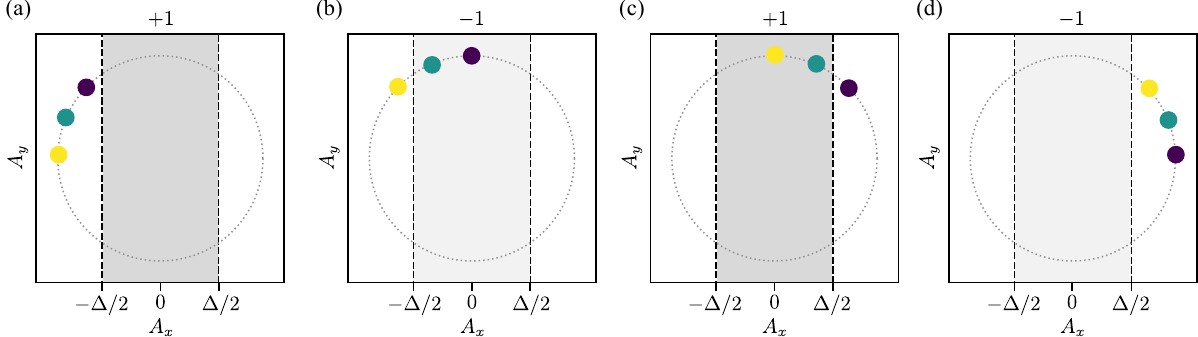}
    \captionsetup{justification=raggedright}
    \caption{\label{fig:process}An example containing three special cases for $K=4$ with $\theta_k = \pi k /4$ for $k \in \{0,1,2,3\}$. The average scores are $1/4$ for the yellow case, $0$ for the green case and $-1/4$ for the purple case. The $+1$ or $-1$ means that for that round, the score is $+1$ or $-2$ when the points are inside the desired region. 
    }
\end{figure*}

\begin{figure*}
    \centering\includegraphics{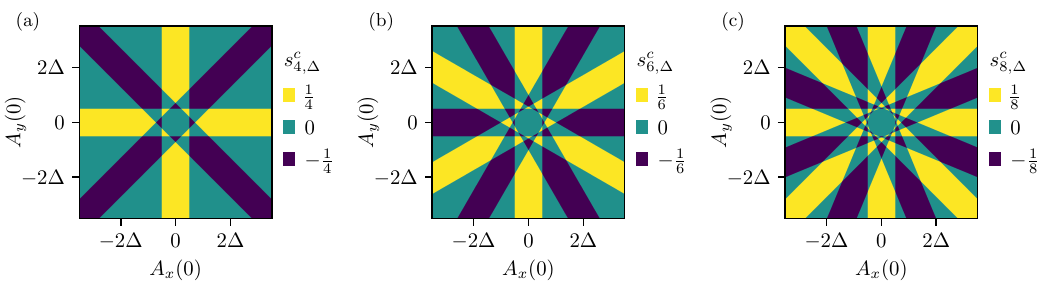}
    \captionsetup{justification=raggedright}
    \caption{\label{fig:classicalScores} The classical score $s_{K,\Delta}^c$ given that the initial state is $(A_x(0),A_y(0))$ for (a) $K = 4$, (b) $K=6$, and (c) $K=8$. We find that $s_{K,\Delta}^c$ takes only the values $0$ and $\pm 1/K$, which implies the classical bound $|s_{K,\Delta}^c| \leq 1/K =: \mathbf{s}_{K}^c$.}
\end{figure*}

We shall call a physical observable $A_x(\theta)$ uniformly precessing with respect to a parameter $\theta$ if it satisfies
\begin{equation}\label{eq:precession-condition}
    \begin{aligned}
    A_x(\theta) &= \cos(\theta) A_x(0) + \sin(\theta) A_y(0),\\
    A_y(\theta) &= \cos(\theta) A_y(0) - \sin(\theta) A_x(0).
    \end{aligned}
\end{equation}
where $A_{\hat{n}}(\theta)$ is the value of $A_{\hat{n}}$ along the angle $\theta$ in the classical case, and an operator in the Heisenberg picture in the quantum case. In earlier works, uniform precession was usually taken to have arisen from the dynamics of the system, as with the position and momentum of a harmonic oscillator with angular frequency $\omega$ at time $t = \theta/\omega$. Alternately, it could similarly arise from spatial rotations, as with the $x$ component of any vector $\vec{A} = (A_x,A_y,A_z)^T$ rotated about the $z$ axis. For the rest of this paper, we shall mostly use the terminology of angles rather than time, although it should not be forgotten that the precession protocol can be applied in the same way to both.

Here, we define the even-parity precession protocol for a uniformly-precessing observable $A_x(\theta)$, a real number $\Delta \geq 0$ separating the score region, and an even integer $K \geq 4$ indicating the number of points divided within the half period, and is is carried out by performing many independent rounds. In each round the following steps occur:

(1)  A value $\theta_k = \pi k /K$ for $k \in \{0,1,\dots,K-1\}$ is randomly chosen.

(2) $A_x(\theta_k)$ is measured at the chosen angle $\theta_k$.

(3) The score $(-1)^k$ is assigned if $|A_x(\theta_k)| \leq \Delta/2$, and the score $0$ is assigned otherwise. This is the main difference from the odd-parity protocols previously defined, where the score was: $1$ when $A_x(\theta_k)>0$, $1/2$ for $A_x(\theta_k)=0$, and $0$ for $A_x(\theta_k) < 0$, for all $k$. 

After sufficiently many rounds, one can estimate the average score within the set $\{\theta_k = \pi k /K : k \in \{0,1,\dots,K-1\}\}$, that is
\begin{equation}
    s_{K,\Delta} := \frac{1}{K} \sum_{k=0}^{K-1} (-1)^k \Pr[ \abs{A_x(\theta_k)} \leq \frac{\Delta}{2} ].
\end{equation}
Here, $\Pr(x)$ is a function such that it returns $1$ when $x$ is true and $0$ otherwise. An example implementation with $K=4$ is shown in Fig.~\ref{fig:process}. 

In classical mechanics, the score $s^c_{K,\Delta}$ of a pure state is fully determined by its initial configuration $(A_x(0),A_y(0))$. The classical scores for $K=2,4,6$ for different initial states are plotted in Fig.~\ref{fig:classicalScores}, where it is observed that $s_{K,\Delta}^c \in \{0,\pm 1/K\}$. Indeed, the formal geometric proof in Appendix~\ref{apd:classical-bound} shows this to be true for every $K$. As general classical states are probability distributions on classical pure states, this in turn implies by convexity that the classical bound is
\begin{equation}
    \abs{s_{K,\Delta}^c} \leq \frac{1}{K} =: \mathbf{s}_{K}^c,
\end{equation}
which is independent of $\Delta$.

\section{\label{sec:quantum}Quantum Violations of the Classical Bound}
We shall now study the violation of the classical bound with two particular examples of uniformly-precessing quantum systems: the continuous variable case of the quantum harmonic oscillator, and the discrete variable case of spin angular momentum.

\subsection{\label{sec:QHO1}Quantum harmonic oscillator}

The quantum harmonic oscillator is specified by the Hamiltonian $H = P^2/(2m) + m\omega^2 X^2/2$, where the position $X$ and momentum $P$ are operators that satisfy the uncertainty relation $\comm{X}{P} = i\hbar \mathbbm{1}$. The evolution of these observables in time is given in the Heisenberg picture as
\begin{equation}
\begin{aligned}
    X(t) &= \cos(\omega t) X + \sin(\omega t) P/(m\omega) \\
    P(t) &= \cos(\omega t) P - \sin(\omega t) m\omega X.
\end{aligned}
\end{equation}
Alternatively, the observable $\sqrt{m\omega/\hbar}\,X(\theta/\omega)$ is the quadrature along the angle $\theta$ associated with homodyne detection in quantum optics \cite{quantum-optics-textbook}. This observable complies with Eq.~\eqref{eq:precession-condition}, so the position of a harmonic oscillator, equivalently the quadrature in optical systems, is uniformly precessing with respect to $\theta$.

\begin{figure}
    \centering\includegraphics{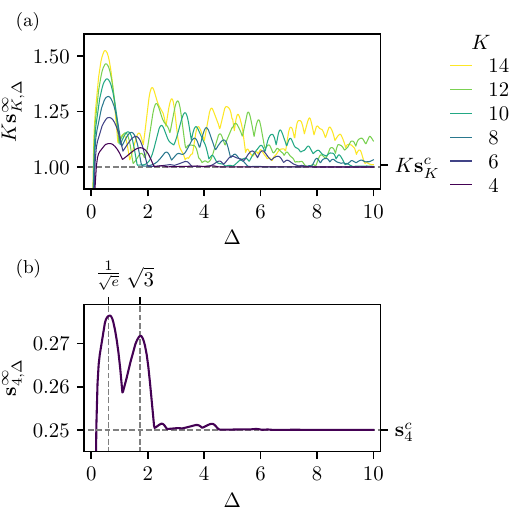}
    \captionsetup{justification=raggedright}
    \caption{\label{fig:cvAgainstDelta}The scaled maximum score $s^{\infty}_{K,\Delta}$ against $\Delta$ for the quantum harmonic oscillator, for (a) selected values of $K$ and (b) a detailed look at $K=4$, where the approximate values of $\Delta$ at the two peaks are also shown. The scores were obtained by constructing the matrix $S^{\infty}_{K,\Delta}$ in the energy eigenbasis of the harmonic oscillator up to some truncation, then numerically calculating its maximum eigenvalue.}
\end{figure}

Hence, we can perform the precession protocol on $X(\theta/\omega)$, where the expected score is given by
\begin{equation}
\begin{aligned}
    s_{K,\Delta}^{\infty} &= \frac{1}{K} \sum_{k=0}^{K-1} (-1)^k \Pr[ \sqrt{\frac{m\omega}{\hbar}}\abs{X\pqty{\frac{\theta_k}{\omega}}} \leq \frac{\Delta}{2}] \\
    &= \tr(\rho S_{K,\Delta}^{\infty}).
\end{aligned}
\end{equation}
Here, the density operator $\rho$ specifies the initial state of the quantum system, and
\begin{equation}\label{eq:define-CV-observable}
    S_{K,\Delta}^{\infty} := \frac{1}{K}\sum_{k=0}^{K-1}(-1)^k e^{i\theta_k \frac{H}{\hbar\omega}} \int_{-\sqrt{\frac{\hbar\Delta^2}{2 m\omega}}}^{\sqrt{\frac{\hbar\Delta^2}{2 m\omega}}}\dd{x}\ketbra{x} e^{-i\theta_k \frac{H}{\hbar \omega}},
\end{equation}
where $\ket{x}$ are the improper eigenvectors of $X$ with eigenvalue $x$. For a quantum harmonic oscillator, the state rotates driven by its Hamiltonian, with the unitary operator $e^{-i\theta_k \frac{H}{\hbar \omega}}$. Compared to the classical case in which the observable is one point on the $x-y$ plane, the state of the quantum harmonic oscillator expands over the whole $x-p$ space. And we define the score operator within the Heisenberg picture, by the rule that the score assigned is zero outside the defined range, so we need to integrate only the effective region. The infinity symbol in the superscript pegs these scores to the quantum harmonic oscillator to differentiate them from the finite-dimensional discrete variable system we will study later.

Obtaining the maximum quantum score $\mathbf{s}^{\infty}_{K,\Delta} := \max_\rho {\tr}(\rho S_{K,\Delta}^{\infty})$ is therefore the same as solving for the largest eigenvalue of $S^{\infty}_{K,\Delta}$. In order to obtain numerical estimates of the quantum score, we construct $S_{K,\Delta}^\infty$ as a matrix in a chosen orthonormal basis up to some truncation and solve for its eigenvalues with standard numerical tools (see Appendix~\ref{apd:qhm-matrix-elements} for details).

The maximum quantum scores $\mathbf{s}_{K,\Delta}^\infty$ obtained using this method are plotted in Fig.~\ref{fig:cvAgainstDelta} against $\Delta$ for some choices of $K$. The dependence of $\mathbf{s}_{K,\Delta}^{\infty}$ on $\Delta$ is highly nontrivial. Nonetheless, it can be eked out from Fig.~\ref{fig:cvAgainstDelta} that (1) the classical bound is saturated or violated in most cases; (2) the first and largest peak occurs at $0 < \Delta < 1$; and (3) the maximum quantum score seems to approach the classical bound for large $\Delta$.

We show in Appendix~\ref{apd:qhm-matches-classical} that $\forall \Delta\in (0,\infty): \mathbf{s}_{K,\Delta}^\infty \geq \mathbf{s}_K^c$, which proves observation (i). We do not have analytical proofs for the other two observations; but we note that performing the precession protocol with $\Delta < 1$ probes features of the quantum state within the $|x| < \sqrt{\hbar/(4m\omega)}$ and $|p| < \sqrt{\hbar m\omega/4}$ region. Thus, observation (ii) points to sub-Planck structure of the Wigner function, i.e., ~features found within an area $\Delta x \Delta p < \hbar$ in phase space \cite{sub-planck}.

\begin{figure*}
    \centering\includegraphics{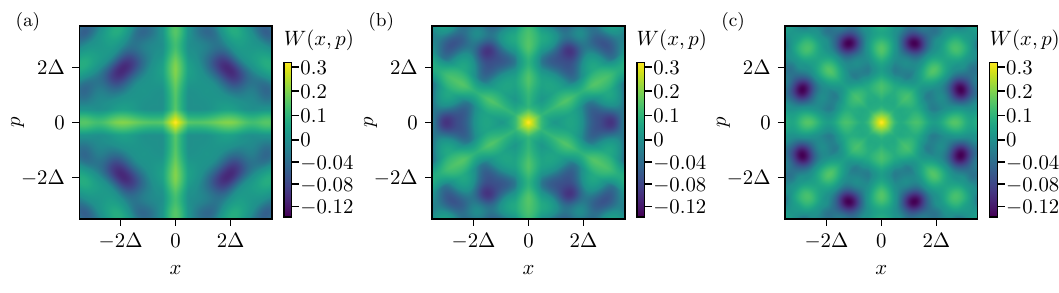}
    \captionsetup{justification=raggedright}
    \caption{\label{fig:wigner}The Wigner function of the quantum harmonic oscillator state that achieves the score $\max_{\Delta}\mathbf{s}_{K,\Delta}^\infty$ for (a) $K=4$, (b) $K=6$, and (c) $K=8$. A comparison with Fig.~\ref{fig:classicalScores} shows that the negativities are concentrated in the region where $\mathbf{s}_{K,\Delta}^c  = -1/K$. Notice also that the states are unchanged under the parity transformation $x \to -x$ and $p \to -p$, which shows that the even-parity precession protocol can detect states that commute with the parity operator.}
\end{figure*}

The Wigner function of the state that maximally violates the precession protocol is plotted in Fig.~\ref{fig:wigner} for $K=4,6,8$ . Comparing the results to Fig.~\ref{fig:classicalScores}, we find that the Wigner negativity is concentrated in the regions where the classical score is $s_{K,\Delta}^c = -1/K$, and $W(x,p)$ take on large positive values when $s_{K,\Delta}^c = 1/K$. The relationship between Wigner negativity and the violation of the classical bound is further quantified in Appendix~\ref{apd:qhm-wigner-negativity}. In particular, we prove the lower bound 
\begin{align}
    K(|s_{K,\Delta}^\infty|-\mathbf{s}_{K}^c) &\leq 2\mathcal{N}_V
\end{align}
on the Wigner negativity volume
\begin{align}
    \mathcal{N}_V := \int\dd{x}\int\dd{p}[|W(x,p)| - W(x,p) ]/2\,,\label{NV}
\end{align}
which quantifies Wigner negativity as a resource \cite{wigner-negativity-volume,non-gaussian-resource-1,non-gaussian-resource-2}.

Finally, from the definition in Eq.~\eqref{eq:define-CV-observable}, it is straightforward to see that $S^{\infty}_{K,\Delta}$ commutes with the parity operator $\Pi$, which has the actions $\Pi X\Pi = -X$ and $\Pi P \Pi = - P$. As such, $S^{\infty}_{K,\Delta}$ and $\Pi$ can be simultaneously diagonalized, which means that there are eigenvectors $\ket{\lambda}$ of $S^{\infty}_{K,\Delta}$ satisfying $\comm{\ketbra{\lambda}}{\Pi} = 0$ that violate the classical bound. A particular example is the states in Fig.~\ref{fig:wigner}. This solves the limitation of the odd-parity precession protocol, which cannot detect states that commute with the parity operator \cite{tsirelson-spin}.

\subsection{Spin angular momentum}
\label{sec:SAM1}
In quantum mechanics, the components of the angular momentum vector $\vec{J} = (J_x,J_y,J_z)$ satisfy the commutation relation $\comm{J_x}{J_y} = i\hbar J_z$, where $\hbar$ is the reduced Planck constant. Let $\ket{j,m_{\hat{n}}}$ denote the simultaneous eigenvector of $\lvert\vec{J}\rvert^2$ and $J_{\hat{n}}$, where $\hat{n} \in \{x,y,z\}$, with associated eigenvalues $\hbar^2 j(j+1)$ and $\hbar m$, respectively. Here, $j$ is a nonnegative integer or half-integer, and $m \in \{-j, -j+1, \dots, j-1, j\}$.

Generally, $\vec{J} = \oplus_j \vec{J}^{(j)}$ is a direct sum of irreducible blocks $\vec{J}^{(j)}$ spanned by the eigenvectors of $\lvert\vec{J}\rvert^2$ with eigenvalue $\hbar^2 j(j+1)$ \cite{angular-momentum-book}. For simplicity, we shall consider each block $\vec{J}^{(j)}$, each with a fixed $j$, separately.

Now, $J_x^{(j)}$ evolving under a rotation generated by the Hamiltonian $H \propto - J_z^{(j)}$ results in the uniformly-precessing observable
\begin{equation}
    J_x^{(j)}(\theta) = e^{-i\theta_k \frac{J_z^{(j)}}{\hbar}} J_x^{(j)} e^{i\theta_k \frac{J_z^{(j)}}{\hbar}} = \cos(\theta)J_x^{(j)} + \sin(\theta)J_y^{(j)}. 
\end{equation}
If the even-parity precession protocol is performed on $J_x^{(j)}/\hbar$, the expected score will be $s_{K,\Delta}^{(j)} = {\tr}(\rho S_{K,\Delta}^{(j)})$ given that the initial state of the system is $\rho$, where
\begin{equation}\label{eq:define-spin-observable}
    S_{K,\Delta}^{(j)} := \frac{1}{K}\sum_{k = 0}^{K-1}(-1)^ke^{-i\theta_k \frac{J_z^{(j)}}{\hbar}} \Pr(\frac{\abs\big{J_x^{(j)}}}{\hbar} \leq \frac{\Delta}{2}) e^{i\theta_k \frac{J_z^{(j)}}{\hbar}},
\end{equation}
with ${\Pr}(|J_x^{(j)}|/\hbar \leq \Delta/2)$ defined in terms of the eigenvectors of $J_x^{(j)}$ as
\begin{equation}\label{eq:define-prob-Jx}
\begin{aligned}
    \Pr(\frac{\abs\big{J_x^{(j)}}}{\hbar} \leq \frac{\Delta}{2}) &= \!\!\sum_{m=-j}^{j}\!\! \Pr(|m| \leq \frac{\Delta}{2}) \ketbra{j,m_x} \\
    &= \!\!\sum_{m=- (\lfloor \frac{\Delta}{2} + j\rfloor - j)}^{\lfloor \frac{\Delta}{2} + j\rfloor - j}\!\!\ketbra{j,m_x} \\
    &= \Pr(\frac{\abs\big{J_x^{(j)}}}{\hbar} \leq \left\lfloor \frac{\Delta}{2}+j \right\rfloor -j).
\end{aligned}
\end{equation}
The maximum quantum score $\mathbf{s}_{K,\Delta}^{(j)} := \max_{\rho} {\tr}(\rho S_{K,\Delta}^{(j)})$ can be found by using commonly-available eigensolvers on the finite-dimensional operator $S_{K,\Delta}^{(j)}$.

\begin{figure}
    \centering
    \includegraphics{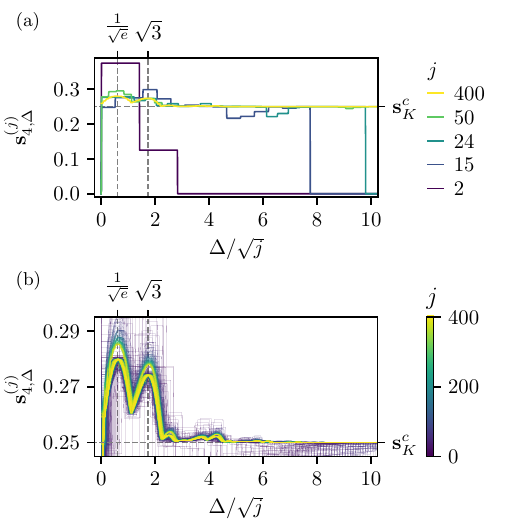}
    
    \captionsetup{justification=raggedright} 
    \caption{\label{fig:spinAgainstJandDelta} The quantum scores $\mathbf{s}_{4,\Delta}^{{j}}$ for spin angular momentum against $\Delta$. (a) Plots for the five specific values $j\in\{2,15,24,50,400\}$ for the whole score range. (b) Superposition of the plots for all integer and half-integer values $0\leq j \leq 400$, zoomed in on the range where the classical bound is violated. The plot of $\mathbf{s}_{4,\Delta}^{{j}}$ appears to converge towards $\mathbf{s}_{4,\Delta}^{\infty}$ (Fig.~\ref{fig:cvAgainstDelta}) for large $j$.}
\end{figure}

These scores are plotted for $K=4$ for several values of $j$ in Fig.~\ref{fig:spinAgainstJandDelta}(a). The stepped behavior is due to the discrete eigenvalues of $J_x$, while the dependence of $\mathbf{s}_{K,\Delta}^{(j)}$ on $\Delta$ is rather complex. However, if we superimpose such plots for many values of $j$ in Fig.~\ref{fig:spinAgainstJandDelta}(b), the traces appear to converge to the harmonic oscillator scores plotted in Fig.~\ref{fig:cvAgainstDelta}. This seems to imply that
\begin{equation}\label{eq:spin-conjectured-limit}
    \lim_{j \to \infty} \mathbf{s}_{K,\Delta/\sqrt{j}}^{(j)} \stackrel{?}{=} \mathbf{s}_{K,\Delta}^{\infty},
\end{equation}
which parallels a similar relationship between the spin angular momentum and harmonic oscillator in the odd-parity precession protocol \cite{tsirelson-spin}, although we have not been able to formally prove Eq.~\eqref{eq:spin-conjectured-limit} in the current work.

\begin{figure}
    \centering
    \includegraphics{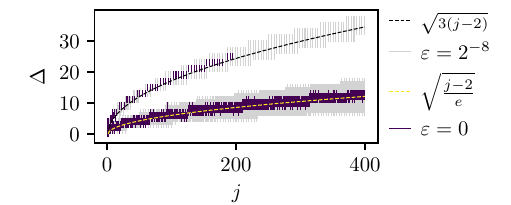}\captionsetup{justification=raggedright} 
    \caption{\label{fig:spinDeltaAgainstJ} $\Delta$ that satisfy $|\mathbf{s}_{4,\Delta}^{(j)}-\max_{\Delta'}\mathbf{s}_{4,\Delta'}^{(j)}| \leq \varepsilon$ for $\varepsilon = 2^{-8}$ (gray line) and for $\varepsilon = 0$ (dark purple line). We find that they cluster around $\sqrt{3(j-2)}$ (dashed black line) and $\sqrt{(j-2)/e}$ (dashed yellow line). Compared to Fig.~\ref{fig:spinAgainstJandDelta}, these two clusters of $\Delta$ correspond to the largest two peaks of $\mathbf{s}_{4,\Delta}^{(j)}$. In particular, the maximum violation for fixed $j$ occurs roughly at $\argmax_\Delta \mathbf{s}_{4,\Delta}^{(j)} \approx \sqrt{(j-2)/e}$.}
\end{figure}

To determine the choice of $\Delta$ that maximizes the quantum violation, we plot the values of $\Delta$ that achieve quantum scores close to the maximum possible score in Fig.~\ref{fig:spinDeltaAgainstJ}. That is, they satisfy $|\mathbf{s}_{4,\Delta}^{(j)}-\max_{\Delta'}\mathbf{s}_{4,\Delta'}^{(j)}| \leq \varepsilon$ for $\varepsilon = 2^{-8}$ (gray line) and for $\varepsilon = 0$ (dark purple line). We find that the optimal values of $\Delta$ tend to cluster around $\sqrt{3(j-2)}$ and $\sqrt{(j-2)/e}$. Cross-referencing them with Figs.~\ref{fig:spinAgainstJandDelta}(b) and Fig.~\ref{fig:cvAgainstDelta}, they correspond to the largest two peaks with $\argmax_\Delta \mathbf{s}_{4,\Delta}^{(j)} \approx \sqrt{(j-2)/e}$, and have the limiting values $\Delta/\sqrt{j} \to 1/\sqrt{e}$ and $\Delta/\sqrt{j} \to \sqrt{3}$. 


\begin{figure}
    \centering\captionsetup{justification=raggedright} 
    \includegraphics{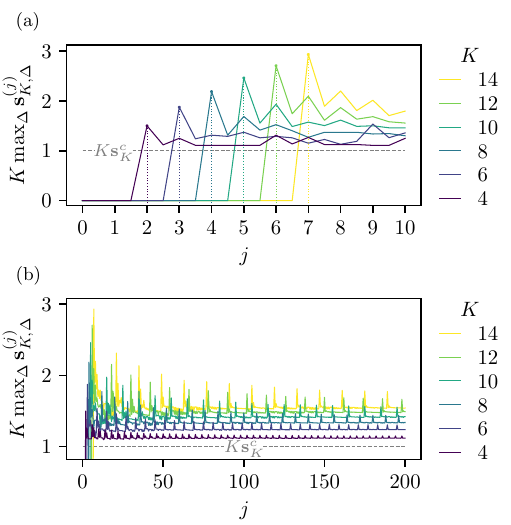}
    \caption{\label{fig:spinAgainstJ}Maximum quantum scores $\max_{\Delta}\mathbf{s}_{K,\Delta}^{(j)}$ against $j$ for (a) $j \leq 10$ and (b) $j \leq 200$. The score is zero whenever $j < K/2$, always peaks at $j=K/2$, violates the classical bound for every $j \geq K/2$, and converges to a limit as $j \to \infty$. We conjecture that the limit is $\max_{\Delta}\mathbf{s}_{K,\Delta}^{\infty}$.}
\end{figure}

For other values of $K$, the maximum quantum scores $\max_{\Delta} \mathbf{s}_{K,\Delta}^{(j)}$ are plotted against $j$ in Fig.~\ref{fig:spinAgainstJ}. Several features can be observed: The score is zero whenever $j < K/2$, the maximum violation occurs at $j=K/2$, the classical bound is violated for every $j \geq K/2$, and the scores appear to converge as $j\to\infty$, which is consistent with the conjectured limit in Eq.~\eqref{eq:spin-conjectured-limit}.

A stronger form of the first observation is proved in Appendix~\ref{apd:spin no violation}, where we show that $\mathbf{s}_{K,\Delta}^{(j<K/2)} = 0$ for every $K$ and $\Delta$. As such, the precession protocol is also a dimension witness, as the violation of the classical bound when performing the protocol on a uniformly-precessing spin certifies that the dimension of the Hilbert space $d$ is lower bounded as $d \geq K+1$.

Meanwhile, the maximum score for $j=K/2$ is analytically worked out in Appendix~\ref{apd:spin case} to be
\begin{equation}
\begin{aligned}
    \mathbf{s}_{K,\Delta}^{(K/2)} &= 2^{-(K-1)}\binom{K-1}{\lfloor \frac{K+\Delta}{2} \rfloor},
\end{aligned}
\end{equation}
which is achieved by the spin cat state
\begin{equation}
\begin{aligned}
    \ket{\mathbf{s}_{K,\Delta}^{(K/2)}} &= \frac{1}{\sqrt{2}}[{
        \ket{j,-j_z} + (-1)^{\lfloor (\Delta+K)/2\rfloor} \ket{j,j_z}
    }].
\end{aligned}
\end{equation}
The maximum value of $\mathbf{s}_{K,\Delta}^{(K/2)}$ for fixed $K$ is given by the central binomial coefficient, which occurs when $\lfloor (K+\Delta)/2 \rfloor = \lceil (K-1)/2 \rceil = K/2$. That is,
\begin{equation}\label{eq:spin-max-score}
\begin{aligned}
    \max_\Delta \mathbf{s}_{K,\Delta}^{(K/2)} &= \mathbf{s}_{K,\Delta=0}^{(j=K/2)} = 2^{-(K-1)}\binom{K-1}{K/2}, \\
    \argmax_{\Delta} \mathbf{s}_{K,\Delta}^{(K/2)} &\in [0,2).
\end{aligned}
\end{equation}
With the closed-form expression in Eq.~\eqref{eq:spin-max-score}, we show in Appendix~\ref{apd:spin case} that $\mathbf{s}_{K,\Delta=0}^{(j=K/2)} > \mathbf{s}_{K}^c$ for all even $K\geq 4$. Therefore, the classical bound of the even-parity precession protocol can be violated by all quantum systems with integral spins $j \geq 2$. This contrasts with the odd-parity precession protocol, in which no violation was found for $j=2$ \cite{tsirelson-spin}.

Finally, while we have taken the spin number $j$ to be fixed in the above discussions, it is easily extendable to the case where the protocol is performed on $\vec{J} = \oplus_{j\in\mathcal{J}} \vec{J}^{(j)}$ for some set $\mathcal{J}$ of spins. Then, the maximum score will be given by $\max_{j \in \mathcal{J}}\mathbf{s}_{K,\Delta}^{(j)}$, as the direct sum structure of $\vec{J}$ will be imposed onto the operator $S_{K,\Delta}$.

\section{\label{sec:entanglement}Witnessing Entanglement}
In this section, we apply the even-parity precession protocol to composite systems, and demonstrate that it is an entanglement witness when performed on certain collective coordinates.

\subsection{Witnessing non-Gaussian entanglement with quadrature measurements}
\label{sec:QHO2}
A system of two identical linearly-coupled harmonic oscillators is governed by the Hamiltonian
\begin{equation}\label{eq:coupled-oscillator-hamiltonian}
    H = \frac{P_1^2 + P_2^2}{2m} + \frac{m\omega^2}{2}\pqty{X_1^2 + X_2^2} - gX_1X_2,
\end{equation}
where $X_n$ and $P_n$ are, respectively, the position and momentum of the $n$th oscillator, $m$ is their mass, $\omega$ is the angular frequency of the trap, and $g$ is the coupling strength.

Equation~\eqref{eq:coupled-oscillator-hamiltonian} can also be rewritten as
\begin{equation}\label{eq:uncoupled-oscillator-hamiltonian}
    H = \frac{P_+^2}{2m} + \frac{m\omega_+^2}{2} X_+^2 + \frac{P_-^2}{2m} + \frac{m\omega_-^2}{2} X_-^2,
\end{equation}
where $X_\pm := (X_2 \pm X_1)/\sqrt{2}$, $P_\pm := (X_2 \pm X_1)/\sqrt{2}$, and $\omega_\pm := \omega \pm g/(2m)$. Here, $X_+$ is the center-of-mass position of the two oscillators, and its evolution in time generated by the Hamiltonian in Eq.~\eqref{eq:uncoupled-oscillator-hamiltonian} is
\begin{equation}
    X_+(t) = \cos(\omega_+ t) X_+ + \sin(\omega_+ t) P_+.
\end{equation}
As such, we can perform the precession protocol on coupled harmonic oscillators using $X_+(\theta/\omega_+)$ as the uniformly-precessing variable. For an initial state $\rho$ of the system, the expected score would be $\mathbf{s}_{K,\Delta}^{\infty} = {\tr}(\rho S_{K,\Delta}^{\infty})$, where $S_{K,\Delta}^{\infty}$ is as defined in Eq.~\eqref{eq:define-CV-observable} with $X_+$ in place of $X$. Now, let $\tr_-$ be the partial trace over the $(X_-,P_-)$ mode. Then, since $S_{K,\Delta}^{\infty}$ is defined only on the center-of-mass mode,
\begin{equation}
    \mathbf{s}_{K,\Delta}^{\infty} = \tr(\rho S_{K,\Delta}^{\infty})
    = \tr_+[ (\tr_-\rho) S_{K,\Delta}^{\infty}].
\end{equation}
With the results of Appendix~\ref{apd:qhm-wigner-negativity}, this means that  $s_{K,\Delta}^\infty \leq \mathbf{s}_{K}^c + 2\mathcal{N}_V/K$, where $\mathcal{N}_V$ is the Wigner negativity volume \eqref{NV} of the state $\tr_-\rho$.

If $\rho$ is separable over the two oscillators, it is known that the Wigner function of $\tr_-\rho$ must be nonnegative via Theorem~2 of Ref.~\cite{CD-entanglement}, which implies that $\mathcal{N}_V = 0$. However, if $\mathcal{N}_V = 0$, then $s_{K,\Delta}^\infty \leq \mathbf{s}_{K}^c$. Taking the contrapositive statement, if the even-parity precession protocol is performed on the center-of-mass position $X_+$, then $s_{K,\Delta}^\infty > \mathbf{s}_{K}^c$ implies that the two oscillators are entangled. Furthermore, since $2\mathcal{N}_V \geq K(s_{K,\Delta}^\infty - \mathbf{s}_{K}^c) > 0$, this criterion detects only non-Gaussian entangled states of the system with negative Wigner functions.

\subsection{Witnessing genuine multipartite entanglement with collective spin measurements}
\label{sec:SAM2}
Let us now consider an ensemble of $N$ particles, where the $n$th particle has spin $j_n$ with angular momentum $\vec{J}^{(j_n)}$, and the total spin $\sum_{n=1}^N j_n$ is an integer. Then, $\vec{J} := \sum_{n} \vec{J}^{(j_n)}$ rotates around the $-z$ axis as
\begin{equation}
    J_x(\theta) = e^{i\theta J_z/\hbar} J_x e^{-i\theta J_z/\hbar} = 
    \cos(\theta) J_x + \sin(\theta) J_y.
\end{equation}
As such, we can perform the precession protocol on the total angular momentum $J_x(\theta) = \sum_{n=1}^N J_x^{(j_n)}(\theta)$ for $K = 2\sum_{n=1}^N j_n$. For initial state $\rho$, the observed score will then be $s_{K,\Delta} = {\tr}(\rho S_{K,\Delta})$ with
\begin{equation}
\begin{aligned}
    S_{K,\Delta} &:= \sum_{k = 0}^{K-1}\frac{(-1)^k}{K}e^{-i\theta_k \frac{J_z}{\hbar}} \Pr(\frac{\abs\big{J_x}}{\hbar} \leq \frac{\Delta}{2}) e^{i\theta_k \frac{J_z}{\hbar}} \\
    = \bigoplus_{\tilde{\jmath}} S_{K,\Delta}^{(\tilde{\jmath})}
    &= 2^{-(K-1)}\binom{K-1}{\lfloor\frac{K}{2+\Delta}\rfloor}\Big(
        \ketbra{\tfrac{K}{2},{\tfrac{K}{2}}_z}{\tfrac{K}{2},-{\tfrac{K}{2}_z}} \\[-1ex]
        &\qquad{}+{}
        \ketbra{\tfrac{K}{2},-{\tfrac{K}{2}}_z}{\tfrac{K}{2},{\tfrac{K}{2}}_z}
    \Big),
\end{aligned}
\end{equation}
where $\tilde{\jmath} \leq \sum_n j_n = K/2$ indexes the irreducible blocks that arise from the usual addition of the angular momentum formula, $S_{K,\Delta}^{(\tilde{\jmath})}$ is as defined in Eq.~\eqref{eq:define-spin-observable}, and the explicit form of $S_{K,\Delta}^{(\tilde{\jmath} \leq K/2)}$ is taken from Appendixes~\ref{apd:spin no violation}~and~\ref{apd:spin case}.

For a state $\rho_{\lnot\text{GME}}$ that is not \emph{genuine multipartite entangled}, that is, if $\rho_{\lnot\text{GME}}$ can be written as a convex combination of states that are separable over some partition of the $N$ particles \cite{GME-coined}, we prove in Appendix~\ref{apd:GME} that
\begin{equation}
    \abs{\tr(\rho_{\lnot\text{GME}} S_{K,\Delta})} \leq \frac{\mathbf{s}_{K,\Delta}}{2} =: \mathbf{s}^{K\text{-sep}}_{K,\Delta}.
\end{equation}
As such, observing $\abs{\tr(\rho S_{K,\Delta})} > \mathbf{s}^{K\text{-sep}}_{K,\Delta}$ implies that $\rho \neq \rho_{\lnot\text{GME}}$. The even-parity precession protocol performed on the total angular momentum of a spin ensemble is therefore a witness of genuine multipartite entanglement.

In the case that all $N$ particles are qubits, the maximally-violating state is 
\begin{equation}
    \ket{\mathbf{s}_{K,\Delta}^{(K/2)}} = \frac{1}{\sqrt{2}}[{ \ket{\uparrow}^{\otimes N} + (-1)^{\lfloor (\Delta+K)/2\rfloor} \ket{\downarrow}^{\otimes N} }],
\end{equation}
where $\ket{\uparrow} := \lvert{\frac{1}{2},{\frac{1}{2}}_z}\rangle$ and $\ket{\downarrow} := \lvert{\frac{1}{2},-{\frac{1}{2}}_z}\rangle$. This is exactly the GHZ state of $N$ qubits, where $N$ is even. 

At the same time, we also show that our protocols can detect entanglement in the presence of global depolarizing noise. The depolarized GHZ state is given by
\begin{equation}
    \rho_{K,\Delta}^{(p_G)} := p_G\frac{\mathbbm{1}}{\tr(\mathbbm{1})}+(1-p_G)\ket{\mathbf{s}_{K,\Delta}^{(K/2)}}\bra{\mathbf{s}_{K,\Delta}^{(K/2)}}.
\end{equation}
The score of this case is $s_{K,\Delta}^{(p_G)} = (1-p_G)\mathbf{s}_{K,\Delta}$, so its entanglement can be detected as long as $p_G<1/2$. Since the GHZ state is genuinely multipartite entangled if and only if $p_G < 1/[2(1-2^{-K})]$ \cite{mixed-GHZ-max-GME}, our protocols detect its entanglement close to the theoretical limit.

Since our witness requires only measurements of the total angular momentum, it  belongs to the family of entanglement witnesses that utilize only collective observables. Past witnesses in this family have been shown to detect Dicke and many-body singlet states \cite{guhne_toth_2009, pezze_review_2018}, but the detection of GHZ states for more than three qubits using such a witness was not partially solved until the odd-parity precession protocol was introduced, which detects GHZ states with an odd number of qubits, was introduced \cite{tsirelson-spin-entanglement}. Our witness therefore completely closes this gap.

\section{Conclusion}
In this work, we introduced a family of even-parity precession protocols that can detect nonclassicality and entanglement by simply measuring a uniformly-precessing observable along different angles, or, equivalently, at one of several different times, in each round. A score is assigned based on whether or not it falls within a region close to the origin, and the average score is calculated after many rounds. If the observed score exceeds the maximum classical score, the nonclassicality of the system is certified.

We computed the classical bound and showed its violation for states of the quantum harmonic oscillators and for all spin systems with $j \geq 2$. Unlike other nonclassicality tests, the protocol does not require simultaneous or sequential measurements. In contrast to previous works on the odd-parity precession protocol, this protocol can also detect states that commute with the parity operator.

Last, we applied the protocols to composite systems. We showed that they can detect non-Gaussian entangled states of two coupled harmonic oscillators and genuine multipartite entanglement of spin ensembles. In particular, GHZ states with an even number of qubits can be detected, thereby completely closing the long-standing problem of detecting GHZ states using collective spin measurements.

\begin{acknowledgments}
This work is supported by the National Research Foundation, Singapore, and A*STAR under its CQT Bridging Grant.  
\end{acknowledgments}

\bibliography{reference}

\appendix
\section{\label{apd:classical-bound}Geometric Proof of the Classical Bound}
In this appendix, we derive the classical bound for the even-parity precession protocol for any $\Delta$ and $K$. We start by parametrizing the pure state of a pair of uniformly-precessing classical variables as
\begin{equation}
    \pmqty{A_x(\theta) \\ A_y(\theta)} = \pmqty{ A_0\cos(\theta + \phi_0) \\ A_0\sin(\theta + \phi_0) },
\end{equation}
where $A_0^2 := A_x^2 + A_y^2$ and $\phi_0 := \arg[A_x(0)+iA_y(0)]$.

For a given initial state, the score is
\begin{equation}
\begin{aligned}
    s_{K,\Delta} &= \frac{1}{K}\sum_{k=0}^{K-1} (-1)^k \Pr[\abs{A_x(\theta_k)} \leq \frac{\Delta}{2}] \\
    &= \frac{1}{K}\sum_{k=0}^{K-1} (-1)^k \Pr[A_0 \abs{\cos(\frac{\pi k}{K} + \phi_0)} \leq \frac{\Delta}{2}].
\end{aligned}
\end{equation}
It can immediately be seen that if $A_0 \leq \Delta/2$, then $A_0 \abs{\cos(\frac{\pi k}{K} + \phi_0)} \leq A_0 \leq \Delta/2$. Hence, since $s_{K,\Delta} \propto \sum_{k=0}^{K-1} (-1)^k = 0$ for all initial states with $A_0 \leq \Delta/2$, we need to consider only the cases when $A_0 > \Delta/2$.

\begin{figure}[ht]
    \centering
    \includegraphics{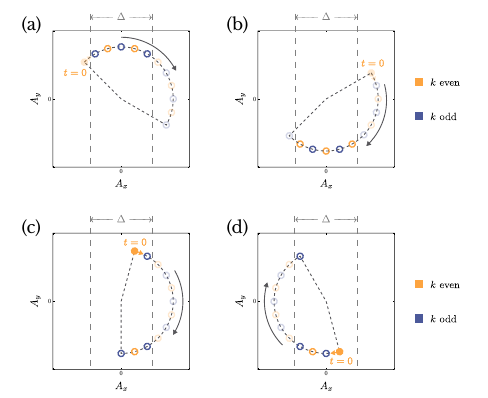}
    \caption{\label{fig:GeometricArgument1}(a)-(d) are four possible trajectories for the even-parity precession protocol when $A_0 > \Delta/2$, shown here for $K=12$.}
\end{figure}

It will be helpful to illustrate this geometrically. There are two main cases, as illustrated in Fig.~\ref{fig:GeometricArgument1}: The trajectory of the particle from $\theta=0$ to $\theta=(K-1)/K$ intersects the $|A_x| \leq \Delta/2$ region (henceforth the $\Delta$-region) only once, like in Figs.~\ref{fig:GeometricArgument1}(a) and ~\ref{fig:GeometricArgument1}(b), or twice, like in Fig.~\ref{fig:GeometricArgument1}(c) and ~\ref{fig:GeometricArgument1}(d).

We can rewrite (A2) as follows:
\begin{equation}
\begin{aligned}
    s_{K,\Delta} &= \pqty{\frac{1}{K}}\Pr[\abs{A_x(\theta_0)} \leq \frac{\Delta}{2}] \\
    &\qquad {}+{} 
    \pqty{ - \frac{1}{K}}
    \Pr[\abs{A_x(\theta_1)} \leq \frac{\Delta}{2}] \\
    &\qquad {}+{} \dots + 
    \pqty{ - \frac{1}{K}}
    \Pr[\abs{A_x(\theta_{K-1})} \leq \frac{\Delta}{2}].
\end{aligned}
\end{equation}
For every term where $k$ is even (e) and within the $\Delta$-region, there is a contribution of $+1/K$ to $s_{K,\Delta}$. Similarly, terms where $k$ is odd (o) and within the $\Delta$-region contribute $-1/K$. Any pair of even and odd points ($e + o$ or $o + e$) within the $\Delta$-region contributes nothing to the score.

\subsection{One intersection}
If there is only a single intersection, there are three possible cases: The extreme points---the leftmost and rightmost points within the $\Delta$-region---are both even, both odd, or different.

Keeping in mind that the points must alternate between even and odd due to the $(-1)^k$ factor, if the extreme points are both even, then the contribution of the points within the $\Delta$-region is
\begin{equation}
\begin{aligned}
    &e + o + e + o + \dots + e + o + e \\
    &=
    \underbrace{(e + o)}_{\!\!\!\!\!\!\text{contributes $0$}\!\!\!\!\!\!} {}+{} (e + o) + \dots + (e + o) + \underbrace{e}_{\hspace{-5em}\text{contributes $+1/K$}\hspace{-5em}} \\
    \implies & s_{K,\Delta} = 1/K.
\end{aligned}
\end{equation}
This is also trivially true when the extreme points are the same, in the sense that there is only one point within the $\Delta$-region.

A similar argument gives $s_{K,\Delta} = -1/K$ when the extreme points are both odd, and $s_{K,\Delta} = 0$ when the extreme points are different. Therefore, the possible classical scores are $s_{K,\Delta}^c \in \{0,\pm 1/K\}$ when there is only one intersection.

\subsection{Two intersections}
With two intersections, one at $A_y > 0$ and another at $A_y < 0$, we can consider the contributions of each intersection individually using the previous technique and then sum them up to obtain the total score. Doing so, it seems that the scores $s_{K,\Delta} = \pm 2/K$ should be possible when the extreme points of both intersections are the same colour. In this section, we show that this can never be the case.

\begin{figure}[h]
    \centering
    \includegraphics{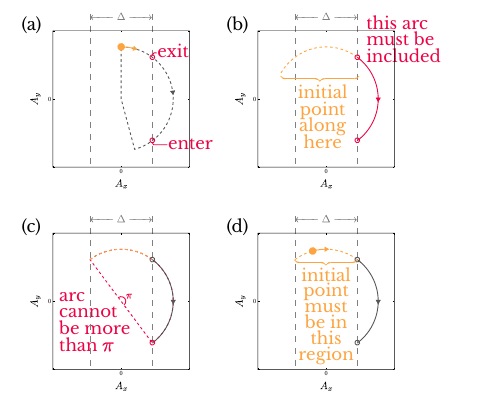}
    \caption{\label{fig:GeometricArgument2}Geometric argument that the initial point must always be within the $\Delta$-region when there are two intersections.}
\end{figure}

We shall again tackle this geometrically. For a trajectory to intersect the $\Delta$-region twice, it has to exit and reenter the region. Since the trajectory is along a circular path while the rectangular region extends from $A_y = -\infty$ to $A_y = \infty$, it must exit and reenter at one of the two edges of the region (the lines $A_x = \Delta/2$ and $A_x = -\Delta/2$). Let us take the case where trajectory crosses $A_x = \Delta/2$, as shown in Fig.~\ref{fig:GeometricArgument2}(a) (the argument proceeds similarly with $A_x = -\Delta/2$). As the precession is clockwise and the center of the circular arc is at the origin, the exit occurs at $A_y > 0$, and the entrance occurs at $A_y < 0$ (conversely, exit at $A_y < 0$ and the entrance at $A_y > 0$ if the trajectory crosses $A_x = -\Delta/2$).

Now, this means that the circular arc that goes from the exit point to the entrance point must be included in the trajectory, and the initial point should occur somewhere before the exit point, as marked in Fig.~\ref{fig:GeometricArgument2}(b).

However, we also know that the full trajectory must be a circular arc that subtends an angle of $\theta_{K-1} - \theta_{0} < \pi$. Hence, as shown in Fig.~\ref{fig:GeometricArgument2}(c), the initial point can be \emph{at most} an angle $\pi$ away from the entrance point (the initial point precedes the exit, which precedes the entrance).

As such, we conclude that the initial point is between an angle $\pi$ away from the entrance, and the exit point. We can see in Fig.~\ref{fig:GeometricArgument2}(d) that this implies that (1) the initial point must be in the $\Delta$-region and (2) the initial point occurs in the $A_y > 0$ plane. Again invoking that the full circular arc must be smaller than a semicircle, the trajectory will not reenter the $\Delta$-region from the left, which means that (3) the initial point is an extreme point of the $\Delta$-region for the intersection at $A_y > 0$. 

These considerations lead us to the conclusion that when the trajectory crosses $A_x = 1/K$, the points within the $\Delta$-region at $A_y > 0$, since one of its extreme points is the even initial point, can contribute only $0$ or $+1/K$ to the total score.

Repeating this argument to find the possible position of the final ($k=K-1$) point, we find that the odd final point (1) must be in the $\Delta$-region, (2) occurs in the $A_y < 0$ plane, and (3) is an extreme point for the intersection at $A_y < 0$. Therefore, when the trajectory crosses $A_x = 1/K$, the points within the $\Delta$-region at $A_y < 0$ can contribute only $0$ or $-1/K$ to the total score.

As such, when the trajectory crosses $A_x = 1/K$, the total score can be at most $s_{K,\Delta}^c = +1/K + 0 = 1/K$ and must be at least $s_{K,\Delta}^c = 0 + (-1/K) = -1/K$. A similar argument gives the same conclusion when the trajectory crosses $A_y = -1/K$, and therefore we have that the classical bound is
\begin{equation}
    \abs{s_{K,\Delta}^c} \leq \frac{1}{K} =: \mathbf{s}_{K,\Delta}^c.
\end{equation}

\section{\label{apd:qhm-matrix-elements}Matrix Elements of \texorpdfstring{$S_{K,\Delta}^\infty$}{Score Observable} for the Quantum Harmonic Oscillator}
In this appendix, we derive the matrix elements of $S_{K,\Delta}^\infty$. We shall first show that $\bra{n}S_{K,\Delta}^\infty\ket{n'}$ is nonzero only when $n-n' = (2l+1)K$, where $n$ and $n'$ are the quantum numbers of the corresponding eigenstates with $H\ket{n} = (n+\frac{1}{2})\hbar\omega$, and $l$ is an integer. 

We begin by defining the averaging operation $E_K^\infty[\bullet]$ as 
\begin{equation}
    E_K^\infty[\bullet] := \frac{1}{K}\sum_{k=0}^{K-1}(-1)^k e^{i\theta_k \frac{H}{\hbar\omega}} [\bullet] e^{-i\theta_k \frac{H}{\hbar \omega}}.
\end{equation}
In terms of this,
\begin{equation}
S_{K,\Delta}^\infty = E_K^\infty\bqty{\int_{-\sqrt{\frac{\hbar\Delta^2}{2 m\omega}}}^{\sqrt{\frac{\hbar\Delta^2}{2 m\omega}}}\dd{x}\ketbra{x}} =: E_K^\infty\bqty{P^x_\Delta},
\end{equation}
where we defined the projector $P^x_\Delta$. Expanding $P^x_\Delta$ in the number basis, we have $P^x_\Delta = \sum_{n,n' = 0}^\infty P^x_{\Delta|n,n'}\ketbra{n}{n'}$ with matrix elements $P^x_{\Delta|n,n'} := \bra{n}P^x_\Delta\ket{n'}$. Plugging this into the expression of $S_{K,\Delta}^\infty$,
\begin{equation}
 E_K^\infty\bqty{P^x_\Delta}
= \frac{1}{K}\sum_{n,n'}P^x_{\Delta|n,n'} \ketbra{n}{n'} \sum_{k=0}^{K-1}(-1)^k e^{-i\theta_k(n-n')}
\end{equation}
With $\theta_k = \pi/K$, we get $\sum_{k=0}^{K-1}(-1)^k e^{-i\theta_k n-n'} = 0$ when $n-n' = 2 l K$ for integer $l$, and $\sum_{k=0}^{K-1}(-1)^k e^{-i\theta_k\frac{n-n'}{\hbar\omega}} = K$ for $n-n' = (2l+1)K$ with an integer $l$. These cover the cases for even $n-n'$ .

When $n-n'$ is odd, we have $e^{-i\pi \frac{H}{\hbar\omega}}\ket{x} = \ket{-x}$ and $\int_{-x_0}^{x_0}\dd{x}\ketbra{x} = \int_{-x_0}^{x_0}\dd{x}\ketbra{-x}$, which tells us that
\begin{eqnarray}
    P^x_{\Delta|n,n'} &=& \bra{n} \bqty{
    \int_{-\sqrt{\frac{\hbar\Delta^2}{2 m\omega}}}^{\sqrt{\frac{\hbar\Delta^2}{2 m\omega}}}\dd{x} e^{-i\pi\frac{H}{\hbar\omega}}\ketbra{-x}{-x} e^{i\pi\frac{H}{\hbar\omega}}
    }\ket{n'}\nonumber \\
    &=& e^{-i\pi(n-n')} P^x_{\Delta|n,n'} = -P^x_{\Delta|n,n'}
\end{eqnarray}
As such, $P^x_{\Delta|n,n'} = 0$ for odd $n-n'$. From the above argument, we conclude that the matrix elements of the operator $S_{K,\Delta}^\infty$ are nonzero only when $n-n' = (2l+1)K$ with an integer $l$.

\begin{widetext}
Using the technique by \citet{Wronskian-trick}, we can also work out $P^x_{\Delta|n,n'}$, which gives us the nonzero matrix elements of $S_{K,\Delta}^\infty$. Therefore,
\begin{equation}\label{eqn: qhm matrix elements}
    \bra{n}S_{K,\Delta}^{\infty}\ket{n'} = \begin{cases}
    \frac{2 e^{-\frac{\Delta^2}{4}}\bqty{
        n'H_{n}(\frac{\Delta}{2})H_{n'-1}(\frac{\Delta}{2}) -
        n H_{n'}(\frac{\Delta}{2})H_{n-1}(\frac{\Delta}{2})
    }}{(n-n')\sqrt{2^{n+n'}n!n'!}} & \text{if $(n-n')\bmod K = 0$ and $(n-n')\bmod (2K) \neq 0$,} \\
    0 & \text{otherwise,}
    \end{cases}
\end{equation}
where $H_n(x) = (-1)^n e^{x^2} \frac{d^n e^{-x^2}}{dx^n}$ are the Hermite polynomials.
\end{widetext} 

\section{\label{apd:qhm-matches-classical}Quantum Score for the Harmonic Oscillator and the Classical Bound}
With a geometric proof similar to that in Appendix~\ref{apd:classical-bound},
it can be shown that initial points $(x,p)$ with $\abs{x}\leq \Delta/2, p>\Delta/[2\sin(\pi/K)]=:p_0$ obtain the score $1/K$. As such, we can lower bound the quantum score in terms of the Wigner function as
\begin{equation}\label{eq:C1}
\begin{aligned}
 s_{K,\Delta}^\infty &\geq \frac{1}{K}\int_{-\frac{\Delta}{2}}^{\frac{\Delta}{2}}\dd{x}\int_{p_0}^{\infty}\dd{p}W(x,p) \\
 &\qquad{}-{}\frac{1}{K}\bqty{1-\int_{-\frac{\Delta}{2}}^{\frac{\Delta}{2}}\dd{x}\int_{p_0}^{\infty}\dd{p}W(x,p)} \\
 &= \frac{1}{K}\bqty{2\int_{-\frac{\Delta}{2}}^{\frac{\Delta}{2}}\dd{x}\int_{p_0}^{\infty}\dd{p}W(x,p)-1}
\end{aligned}
\end{equation}
Given the state $W(x,p) \propto e^{-\sigma^2 x^2} e^{-(p-p_0-\Delta\sigma^2/2)^2/\sigma^2}$, the lower bound approaches $1/K$ as $\sigma \to 0$, which implies that the maximum score of the quantum harmonic oscillator is at least the classical bound.
\section{\label{apd:qhm-wigner-negativity}Relationship Between Quantum Violation and Wigner negativity}

    The score achieved by a state given by $W(x,p)$ can be calculated as
    \begin{equation}
    \begin{aligned}
        s_{K,\Delta}^\infty &= \frac{1}{K}\sum_{k=1}^\infty (-1)^k \int\dd{x}\int\dd{p} \Theta(|x|\leq \Delta/2) W(x,p;t_k), \\
        &= \frac{1}{K}\sum_{k=1}^\infty (-1)^k \int\dd{x}\int\dd{p} \Theta(|x|\leq \Delta/2) \\ &\quad W(\cos(t_k) x - \sin(t_k) p,\cos(t_k) p + \sin(t_k) x;t=0), \\
        &= \int\dd{x}\int\dd{p} \frac{1}{K}\sum_{k=1}^\infty (-1)^k \\ &\quad \Theta[|\cos(t_k)x+\sin(t_k)p|\leq \Delta/2]W(x,p), \\
        &= \mathbf{s}_{K,\Delta}^c \int\dd{x}\int\dd{p} \bqty{\Theta_+(x,p) - \Theta_-(x,p)} W(x,p),
    \end{aligned}
    \end{equation}
    where $\Theta_{\pm}(x,p)$ take the value $1$ in the regions where the classical score is $\pm\mathbf{s}_{K,\Delta}^c$ and $0$ otherwise. The regions are mutually exclusive, so $|\Theta_{+}(x,p)-\Theta_{-}(x,p)| \leq 1$. Writing $W(x,p) = W^{+}(x,p) - W^-(x,p)$, with $W^{\pm}(x,p) := [|W(x,p)| \pm W(x,p)]/2 \geq 0$,
    \begin{equation}\label{eq:D3}
    \begin{aligned}
        \frac{s_{K,\Delta}^\infty}{\mathbf{s}_{K,\Delta}^c } &=  \int\dd{x}\int\dd{p} \bqty{\Theta_+(x,p) - \Theta_-(x,p)} W(x,p), \\
        s_{K,\Delta}^\infty K &= \int\dd{x}\int\dd{p} W(x,p) \\ &+ \int\dd{x}\int\dd{p} \underbrace{\bqty{1 +  \Theta_-(x,p) - \Theta_+(x,p)}}_{\geq 0,\leq 2} \bqty{-W(x,p)} \\
        &\leq \int\dd{x}\int\dd{p} W(x,p) + 2\int\dd{x}\int\dd{p}  W^-(x,p) \\
        &=  2\mathcal{N}_V + 1,
    \end{aligned}
    \end{equation}
where $\mathcal{N}_V := \iint\dd{x}\dd{p} W^-(x,p)$ is the Wigner negativity volume.
    
Rearranging (\ref{eq:D3}) and noting that $\mathbf{s}_{K,\Delta}^c = 1/K$, we get
\begin{equation}\label{eq:D1}
    K(|s_{K,\Delta}^\infty|-\mathbf{s}_{K,\Delta}^c) \leq 2\mathcal{N}_V.
\end{equation}

\section{\label{apd:spin no violation}Scores when \texorpdfstring{$j < K/2$}{spin is less than half the number of probing times}}
In this appendix, we derive the matrix elements of $S_{K,\Delta}^{(j)}$. As the steps are similar to the quantum harmonic oscillator, we provide only a sketch of the proofs: Readers can refer to Appendix~\ref{apd:qhm-matrix-elements} for the details.

We again define
\begin{equation}
    E_K^{(j)}[\bullet] := \frac{1}{K}\sum_{k = 0}^{K-1}(-1)^ke^{-i\theta_k \frac{J_z^{(j)}}{\hbar}} [\bullet] e^{i\theta_k \frac{J_z^{(j)}}{\hbar}},
\end{equation}
where we have $S_{K,\Delta}^{(j)} = E_K^{(j)}[P^{(j)}_\Delta]$ in terms of the projector
\begin{equation}
\begin{aligned}
P^{(j)}_\Delta &:= \Pr(\frac{|J_x^{(j)}|}{\hbar} \leq \frac{\Delta}{2}) \\
&=\hspace{-1em}\sum_{m,m' = -j}^j P^{(j)}_{\Delta|m,m'}\ketbra{j,m_z}{j,m_z'}.
\end{aligned}
\end{equation}
As before, $\bra{j,m_z}S_{K,\Delta}^{(j)}\ket{j,m_z'} = 0$ when $m-m' = 2lK$ for integer $l$, and $\bra{j,m_z}S_{K,\Delta}^{(j)}\ket{j,m_z'} = P^{(j)}_{\Delta|m,m'}$ when $m-m' = (2l+1)K$ for integer $l$. These cover the cases for even $m-m'$.


When $m-m'$ is odd, $e^{-i\pi J_z/\hbar}J_x e^{i \pi J_z/\hbar} = -J_x$ implies that $e^{-i\pi J_z/\hbar} P_{\Delta}^{(j)} e^{i \pi J_z/\hbar} = P_{\Delta}^{(j)}$, where
\begin{equation}
\begin{aligned}
    P_{\Delta|m,m'}^{(j)} &= \bra{j,m_z} e^{-i\pi \frac{J_z}{\hbar}} P_{\Delta}^{(j)} e^{i \pi \frac{J_z}{\hbar}} \ket{j,m'_z} \\
    &= e^{-i\pi(m-m')} P^{(j)}_{\Delta|m,m'} = -P^{(j)}_{\Delta|m,m'}.
\end{aligned}
\end{equation}
As such, $P_{\Delta|m,m'}^{(j)}=0$ for odd $m-m'$.

From the derivation above, the matrix elements of $S_{K,\Delta}^{(j)}$, written in the basis of $J_z$, are nonzero only when $m-m' = (2l+1)K$ for integer $l$. When $j<K/2$, $|m-m'| \leq K$ implies that $m-m' \neq (2l+1)K$ for all $m$ and $m'$, which implies that the matrix is a zero matrix. Therefore, $s_{K,\Delta}^{(j < K/2)} = 0$. 

\section{\label{apd:spin case}Scores when \texorpdfstring{$j = K/2$}{spin is the same as the number of probing times}}

When $j = K/2$, the only nonzero matrix elements of $S_{K,\Delta}^{(j)}$ are $\langle{\frac{K}{2},\pm\frac{K}{2}_z}\rvert S_{K,\Delta}^{(j)}\lvert{\frac{K}{2},\mp\frac{K}{2}_z}\rangle$. The eigenvalue of $S_{K,\Delta}^{(j=K/2)}$ can be solved with standard techniques, which gives the maximum score 
\begin{equation}
\begin{aligned}
\mathbf{s}_{K,\Delta}^{(j)} &= \abs{\bra{\tfrac{K}{2},\tfrac{K}{2}_z} \Pr(\frac{\abs\big{J_x^{(j)}}}{\hbar} \leq \frac{\Delta}{2}) \ket{\tfrac{K}{2},-\tfrac{K}{2}_z}} \\
&= \abs{\sum_{m=- (\lfloor \tfrac{\Delta}{2} + \tfrac{K}{2}\rfloor - \frac{K}{2})}^{\lfloor \tfrac{\Delta}{2} + \tfrac{K}{2}\rfloor - \frac{K}{2}}\hspace{-2em}\braket{\tfrac{K}{2},\tfrac{K}{2}_z}{\tfrac{K}{2},m_x}\!\braket{\tfrac{K}{2},m_x}{\tfrac{K}{2},-\tfrac{K}{2}_z} }
\end{aligned}
\end{equation}
The quantities $\braket{j,m_z}{j,m_x'}$, which are elements of the so-called ``Wigner $d$-matrix'', have the known relations \cite{angular-momentum-book}
\begin{equation}
\begin{aligned}
    \braket{j,m_z'}{j,m_x} &= (-1)^{m-m'}\braket{j,m_z}{j,m_x'} \\
    &= (-1)^{j+m}\braket{j,-m_z'}{j,m_x},
\end{aligned}
\end{equation}
and
\begin{equation}
    \braket{j,j_z}{j,m_x}= 2^{-j}\sqrt{\binom{2j}{j-m}}.
\end{equation}
With these relations, we obtain
\begin{equation}
    \bra{\tfrac{K}{2},\tfrac{K}{2}_z} \ket{\tfrac{K}{2},m_x},
    = (-1)^{\frac{K}{2}-m}\, 2^{-\frac{K}{2}}\sqrt{\binom{K}{\frac{K}{2}-m}}
\end{equation}
\begin{equation}
    \bra{\tfrac{K}{2},m_x}\ket{\tfrac{K}{2},-\tfrac{K}{2}_z}
    =  2^{-\frac{K}{2}}\sqrt{\binom{K}{\frac{K}{2}-m}}.
\end{equation}
We can calculate the maximum score to be
\begin{equation}
\begin{aligned}
\mathbf{s}_{K,\Delta}^{(j)} 
&= \abs{\sum_{m=-(\lfloor \tfrac{\Delta}{2} + \tfrac{K}{2}\rfloor - \frac{K}{2})}^{\lfloor \tfrac{\Delta}{2} + \tfrac{K}{2}\rfloor - \frac{K}{2}}\hspace{-2em}
(-1)^{\frac{K}{2}-m}\, 2^{-K}\binom{K}{\frac{K}{2}-m}
} \\
&= 2^{-(K-1)}\binom{K-1}{\lfloor\frac{K+\Delta}{2}\rfloor}.
\end{aligned}
\end{equation}
This quantity is maximized for the central binomial coefficient $\max_\Delta \binom{K-1}{\lfloor\frac{K+\Delta}{2}\rfloor} = \binom{K-1}{\frac{K}{2}}$. Using mathematical induction, we can also prove that $\max_\Delta\mathbf{s}_{K,\Delta}^{(j=K/2)} = 2^{-K+1}\binom{K-1}{\frac{K}{2}} > 1/K = \mathbf{s}_{K}^c$ when $K\geq 4$, which means that spin angular momentum systems can violate the classical bound for all $K\geq 4$.

\section{\label{apd:GME}Proof of the Separable Bound for Spin Ensembles}
Partition a system into two, and write a pure separable state with respect to this bipartition as $\ket{\psi} = \ket{\psi_1}\otimes\ket{\psi_2}$, where $\ket{\psi_{1,2}}$ denote the states corresponding to each subsystem. For such a state, the observed score is
\begin{eqnarray}
    S_{\mathrm{sep}} &=& \bra{\psi_1}\otimes\bra{\psi_2}S_{K,\Delta}\ket{\psi_1}\otimes\ket{\psi_2} \nonumber\\
    &=& \bra{\psi_1}\Tr_2[S_{K,\Delta}(I_1\otimes\ket{\psi_2}\bra{\psi_2})]\ket{\psi_1}.
\end{eqnarray}
Thus, the task at hand is to determine the maximum eigenvalue of the operator $\Tr_2[S_{K,\Delta}(I_1\otimes\ket{\psi_2}\bra{\psi_2})]$.

Here, we can define $\ket{0} = \otimes\ket{j_n,-j_n}$ and $\ket{1}=\otimes\ket{j_n,j_n}$, where $\ket{j_n,\pm j_n}$ are eigenstates of $J_z$. Then we can write $I_1= \ket{0}_1\bra{0}_1+\ket{1}_1\bra{1}_1 + \sum_k\ket{k}_1\bra{k}_1$ and $\ket{\psi_2} = C_1\ket{0}_2+\ket{1}_2+\sum_lC_l\ket{l}_2$. From the expression of $S_{K,\Delta}$, it is not difficult to get 
\begin{equation}
    S_{K,\Delta}(\ket{k}_1\bra{k}_1\otimes\ket{l}_2\bra{l}_2) = 0
\end{equation}
So finally we can simplify the expression as 
\begin{equation}
\begin{split}
    &\Tr_2[S_{K,\Delta}(I_1\otimes\ket{\psi_2}\bra{\psi_2})] \\
    &= 2^{-K+1}\binom{K-1}{\lfloor \frac{K+\Delta}{2}\rfloor}  (C_1C_2^*\ket{1}_1\bra{0}_1+C_1^*C_2\ket{0}_1\bra{1}_1)
\end{split}
\end{equation} 
We get eigenvalue $\lambda_\pm = \pm \abs{C_1}\abs{C_2}$. With $\abs{C_1}^2+\abs{C_2}^2\leq 1$, the maximum score  is $\textbf{s}_{K,\Delta} = 2^{-K}\binom{K-1}{\lfloor \frac{K+\Delta}{2}\rfloor}$.
\end{document}